\documentclass[a4paper,twocolumn,3p,preprint]{elsarticle}

\usepackage{lineno,hyperref}
\usepackage[usenames,dvipsnames,svgnames,table]{xcolor}
\modulolinenumbers[5]

\journal{Physica E}

\newcommand{\wmk}{ W m$^{-1}$K$^{-1}$ }
\newcommand{\nm}{ N m$^{-1}$ }

\begin{document}

\title{Anomalous strain effect on the thermal conductivity of borophene: a reactive molecular dynamics study.}
\author[buw]{Bohayra Mortazavi\corref{cor1}}
\ead{bohayra.mortazavi@gmail.com}

\author[nam]{Minh-Quy Le}

\author[tong]{Timon Rabczuk}

\author[ufrn]{Luiz Felipe C. Pereira\corref{cor1}}
\ead{pereira@fisica.ufrn.br}

\address[buw]{Institute of Structural Mechanics, Bauhaus-Universit\"at Weimar, Marienstr. 15, D-99423 Weimar, Germany}
\address[nam]{Department of Mechanics of Materials and Structures, School of Mechanical Engineering, Hanoi University of Science and Technology, No. 1, Dai Co Viet Road, Hanoi, Vietnam}
\address[tong]{College of Civil Engineering, Department of Geotechnical Engineering, Tongji University, Shanghai, China}
\address[ufrn]{Departamento de F\'{\i}sica, Universidade Federal do Rio Grande do Norte, Natal, 59078-970, Brazil}

\cortext[cor1]{Corresponding authors}

\date{\today}

\begin{abstract}
Borophene, an atomically thin, corrugated, crystalline two-dimensional boron sheet, has been recently synthesized. 
Here we investigate mechanical properties and lattice thermal conductivity of borophene using reactive molecular dynamics simulations. 
We performed uniaxial tensile strain simulations at room temperature along in-plane directions, and found 2D elastic moduli of 188 \nm and 403 \nm along zigzag and armchair directions, respectively. 
This anisotropy is attributed to the buckling of the borophene structure along the zigzag direction. 
We also performed non-equilibrium molecular dynamics to calculate the lattice thermal conductivity. 
Considering its size-dependence, we predict room-temperature lattice thermal conductivities of $75.9 \pm 5.0$ \wmk and $147 \pm 7.3$ \wmk, respectively, and estimate effective phonon mean free paths of $16.7 \pm 1.7$ nm and $21.4 \pm 1.0$ nm for the zigzag and armchair directions. 
In this case, the anisotropy is attributed to differences in the density of states of low-frequency phonons, with lower group velocities and possibly shorten phonon lifetimes along the zigzag direction. 
{We also observe that when borophene is strained along the armchair direction there is a significant increase in thermal conductivity along that direction. Meanwhile, when the sample is strained along the zigzag direction there is a much smaller increase in thermal conductivity along that direction. For a strain of 8\% along the armchair direction the thermal conductivity increases by a factor of 3.5 (~250\%), whereas for the same amount of strain along the zigzag direction the increase is only by a factor of 1.2 (~20\%).}
Our predictions are in agreement with recent first principles results, at a fraction of the computational cost. 
The simulations shall serve as a guide for experiments concerning mechanical and thermal properties of borophene and related 2D materials.
\end{abstract}

\maketitle

\section{Introduction}

Boron presents a wealth of possible two-dimensional (2D) allotropes, and boron sheets exhibit various structural polymorphs containing mostly hexagonal and triangular lattices \cite{Tang2007,Penev2012,Zhou2014a,Zhang2015c,Mannix2015,Feng2016a}. Low-buckled borophene sheets in the form of triangular lattices have been predicted years ago \cite{Tang2007, Kunstmann2006}, and have recently been synthesized on Ag substrates \cite{Mannix2015}. Electronic, magnetic, mechanical and optical properties of triangular boron sheets have recently been investigated by first-principles calculations \cite{Mortazavi2016b,Lherbier2016}. These 2D boron sheets, so-called borophene, were even predicted to exhibit superconductive behavior \cite{Penev2016}. Recent first-principles calculations confirmed that borophene sheets can serve as an ideal anode electrode material with high electrochemical performance for Mg, Na, and Li ion batteries, which outperform other 2D materials \cite{Mortazavi2016c}. These outstanding physical properties of borophene place it as a direct rival for graphene in a serie of applications \cite{Novoselov2004,Geim2007,Wehling2008}. Nevertheless, in spite of the promising applications of borophene, studies related to its thermal and mechanical properties at finite temperatures are still very limited. In particular, the thermal conductivity of borophene films remained almost unexplored. The mechanical properties of flat boron sheets with different vacancy ratios have recently been investigated via classical molecular dynamics (MD) simulations \cite{Le2016}, where it was shown that their mechanical properties depend significantly on atomic structures, loading direction, and temperature.

Motivated by the most recent experimental advances in the fabrication of corrugated triangular borophene sheets and their wide potential applications \cite{Mannix2015}, we believe it is fundamental to provide a comprehensive understanding of the thermal and mechanical properties of borophene sheets. Therefore, in the present work, we study the lattice thermal conductivity and mechanical properties of corrugated borophene at room temperature by performing extensive reactive molecular dynamics simulations. In general, molecular dynamics simulations are a powerful tool to predict and understand the physical properties of novel materials \cite{Mortazavi2015a,Abadi2016,Mortazavi2016a,Pereira2016}, at a fraction of the computational cost of first-principles calculations. In the present investigation we report on the direction dependent elastic modulus, stress-strain response and phonon thermal conductivity of corrugated borophene at room temperature. Our reactive MD simulations show that corrugated borophene sheets present highly anisotropic thermal and mechanical properties, and can guide future studies on the thermal and mechanical properties of borophene films. 

\section{Methods: Molecular dynamics modelling}

All molecular dynamics simulations were carried out with LAMMPS \cite{Plimpton1995}, where the atomic equations of motion were time-integrated with the velocity Verlet algorithm \cite{Verlet1967}. The ReaxFF \cite{Weismiller2010} potential was  used to model the atomic interactions in borophene sheets. Periodic boundary conditions were applied in the planar directions to remove the effects of free atoms at the edges. In this way, we studied infinite borophene sheets and not 2D boron nanoribbons. In the out of plane direction (z-direction), we defined a $20$ \AA~ vacuum by  fixing  the simulation box size along that direction. All MD simulations in this work were carried out at room temperature ($300$ K). 

First we investigated the mechanical properties of single-layer boron sheets. In this case, the equations of motion were integrated with a small time step of $0.25$ fs. Before applying the loading conditions, the structures were relaxed to zero stress using a Nos\'e-Hoover barostat and thermostat (NPT) for $1.25$ ps. Uniaxial tension in the armchair and zigzag directions (notations are indicated in figure \ref{fig1}) were imposed by applying a constant engineering strain rate of $2.5 \times 10^8$ $s^{-1}$, one direction at a time. Because we are dealing with single-layer boron sheets, the stress perpendicular to the sheet (in the z-direction) is zero. Therefore, in order to guarantee uniaxial stress conditions, zero stress condition in the edge parallel to the tensile direction was achieved by relaxing the size of the simulation box in the direction perpendicular to the tensile direction using a barostat set to zero pressure. The macroscopic stress tensor is given by the virial theorem \cite{Marc1985,Zimmerman2004}:
\begin{equation}
\mathbf{\sigma} = \frac{1}{V} \sum_{a \in V} \left[ -m^{a} \mathbf{v}^a \otimes \mathbf{v}^a + \frac{1}{2} \sum_{a \ne b} \mathbf{r}^{ab} \otimes \mathbf{f}^{ab} \right]
\end{equation}
here,  $m^a$ and $\mathbf{v}^a$ are the mass and the velocity vector of atom $a$, respectively. The symbol $\otimes$ denotes the tensor product of two vectors. $\mathbf{r}^{a}$ denotes the position of atom $a$.  $\mathbf{r}^{ab} = \mathbf{r}^{b} - \mathbf{r}^{a}$ is the distance vector between  atoms $a$ and $b$. $\mathbf{f}^{ab}$ is the force on atom $a$ due to atom $b$. $V$ is the volume of the structure. For boron sheets we defined $V=A \times t$, where $A$ is the surface area of the sheet, and $t$ is the sheet nominal thickness. In all simulations we assumed a nominal thickness of $2.9$ \AA~ for single-layer borophene sheets \cite{Mannix2015}.

The non-equilibrium molecular dynamics (NEMD) method was employed to study the phonon thermal conductivity of borophene. In this method, simulations were performed for borophene samples of increasing length with a fixed nominal width of $6.6$ nm. The phononic thermal conductivity of an infinite borophene sheet, as well as an effective phonon mean free path were extracted from the size dependence of the thermal conductivity. Due to the non-equilibrium conditions in these simulations, it was necessary to use a smaller integration time step, namely a time increment of $0.2$ fs. After obtaining the equilibrated structures at $300$ K, we fixed several rows of boron atoms at the two opposing ends of the simulation sample. Next, the simulation box (excluding the fixed atoms) was divided into $22$ slabs along the sample length, and a $20$ K temperature difference between the first and 22nd slabs was imposed in the system. To this aim, the temperature in these two slabs was controlled at the desired values ($310$ K and $290$ K) by independent Nos\'e-Hoover thermostats, while the remaining slabs were not connected to any thermostats or heat baths, effectively evolving in a microcanonical ensemble. Therefore, a constant heat flux was imposed in the system by continuously adding or removing energy in the thermostated slabs at a rate $dq/dt$. The heat flux along a cartesian direction can then be obtained from: 
\begin{equation}
J_x=\frac{1}{A} \frac{dq}{dt}.
\end{equation}
Here, $A$ is the cross sectional area of the borophene sheets, once again assuming a nominal thickness of $2.9$ \AA~ \cite{Mannix2015}. The temperature of each slab is taken and the average kinetic energy of the atoms in the slab, and converted via the equipartition theorem. After a transient period, the system reaches a steady-state heat transfer condition, and a constant temperature gradient, $dT/dx$, is established along the sample length. The NEMD simulations were performed for at least $6$ ns in the steady state regime, during which the heat current and the temperature gradient were time-averaged. Finally, the thermal conductivity was obtained from Fourier's law:
\begin{equation}
\kappa = \frac{J_x}{dT/dx}.
\end{equation}

In order to calculate the phonon density of states, we simulated a borophene sheet at room temperature under microcanonical conditions for $100$ ps, during which we record the trajectories and velocities. We then post-process the trajectories to calculate the DOS from the Fourier transform of the normalized velocity autocorrelation function, such that:
\begin{equation}
DOS(\omega) = \int_0^{\infty} \frac{ \langle \mathbf{v}(t) \cdot \mathbf{v}(0) \rangle}{\langle \mathbf{v}(0) \cdot \mathbf{v}(0) \rangle} \exp(-i \omega t) dt
\end{equation}
where $\omega$ is the angular frequency and $\mathbf{v}$ is the atomic velocity.

\section{Results}

\begin{figure}[htb]
\centering
\includegraphics[width=\linewidth]{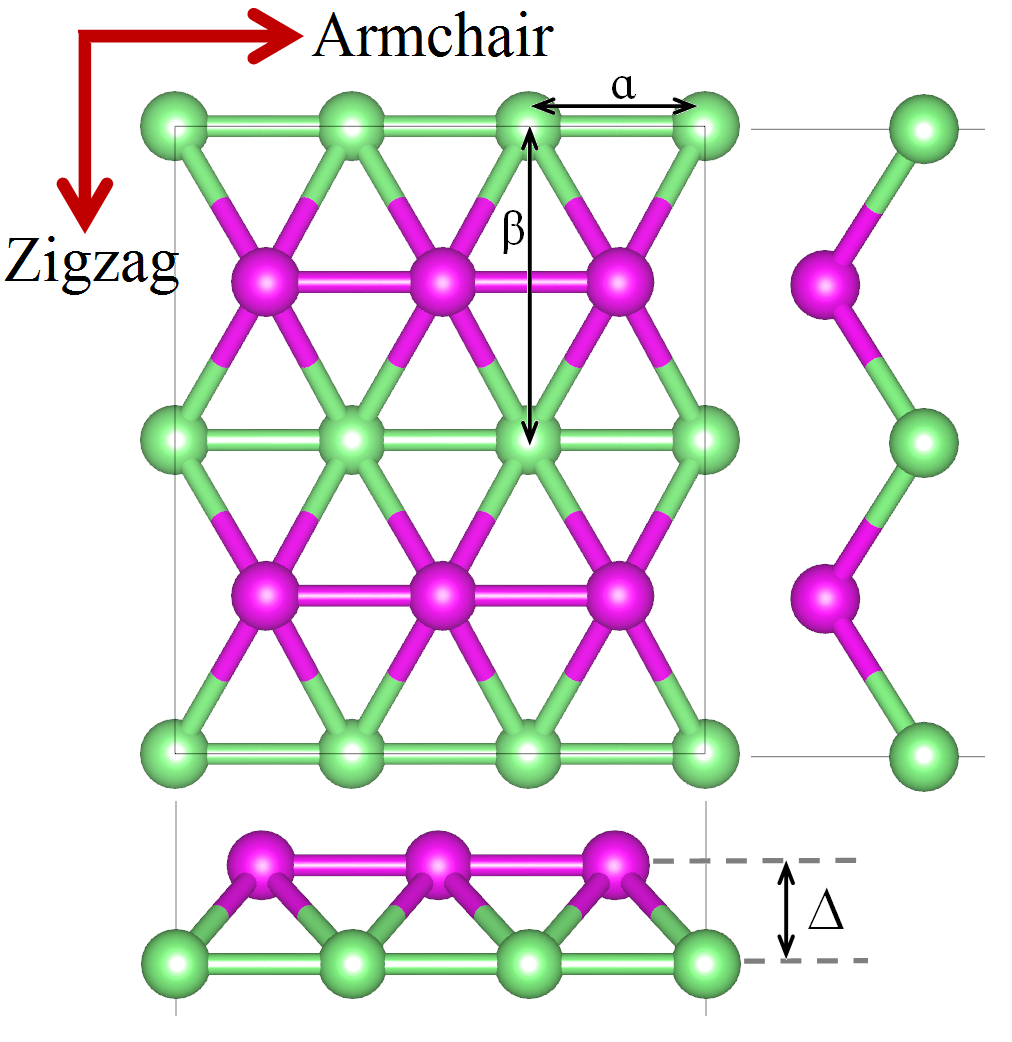}
\caption{Top and side views of the atomic structure of borophene. Two colors, green and purple, are used to indicate the height of the atoms. Lattice constants $\alpha$ and $\beta$, as well as the bucking height $\Delta$ are indicated. We studied mechanical and thermal transport properties along the in-plane directions, indicated as armchair and zigzag in analogy with graphene.}
\label{fig1}
\end{figure}

The atomic structure of corrugated borophene is shown in figure \ref{fig1}. The structure of borophene can be defined by introducing the $\alpha$ and $\beta$ lattice constants and the bucking height, $\Delta$  indicated in figure \ref{fig1}. Based on the ReaxFF results for the relaxed structure, the borophene $\alpha$, $\beta$ and $\Delta$ lattice constants were $2.1$ \AA, $3.18$ \AA~ and $0.76$ \AA, respectively. We note that according to first-principles DFT calculations \cite{Mannix2015} the $\alpha$ and $\beta$ lattice constants of borophene were predicted to be $1.7$ \AA~ and $2.9$ \AA, respectively, while $\Delta$ was calculated at $0.80$ \AA. This indicates that ReaxFF does not accurately predict the in-plane lattice parameters of borophene, at least if the first principles calculations are to be taken as accurate. Nonetheless, we will show in what follows that our simulation results for the elastic modulus and the lattice thermal conductivity are in excellent agreement with first-principles calculations.  

\begin{figure}[htb]
\centering
\includegraphics[width=\linewidth]{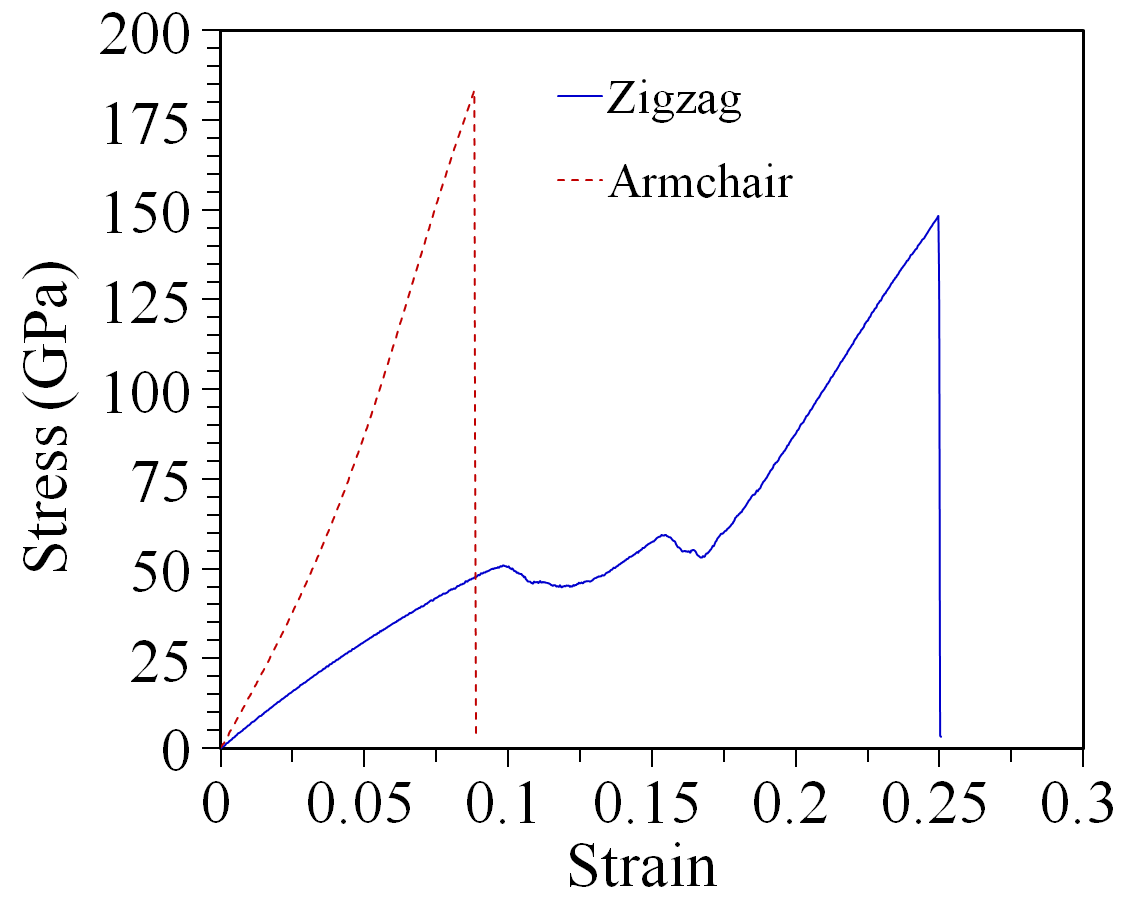}
\caption{Uniaxial stress-strain curves of the borophene sheet at $300$ K. Stress values were calculated assuming a nominal  thickness of $2.9$ \AA~ for single-layer borophene films.}
\label{fig2}
\end{figure}

Figure \ref{fig2} plots the uniaxial stress-strain curves of the borophene sheet at $300$ K along armchair and zigzag directions. Stress values were calculated assuming a nominal thickness of $2.9$ \AA~ for single-layer borophene films. 
The stress-strain curves include an initial linear region which is followed by a nonlinear trend up to a peak value,  the sample tensile strength.
As the strain is increased further, the stress drops suddenly, which is a typical indication of a brittle fracture mechanism. 
Anisotropic tensile strengths of $147$ GPa and $182$ GPa are obtained when the borophene membrane is stretched along zigzag and armchair directions, respectively, as shown in figure \ref{fig2}.
Both values are larger than a recent first principles prediction \cite{Mortazavi2016b}.
Meanwhile, the strain at the tensile strength, know as the fracture strain is estimated at $25$\% for the zigzag direction and $8.8$\% for the armchair direction, while the corresponding first principles values are $14.5$\% and $10.5$\% \cite{Mortazavi2016b}.

The calculated 2D Young's modulus of borophene are $188$ \nm and $403$ \nm in the zigzag and armchair directions, respectively. Remarkably, these values are within $10$\% of the results predicted by first-principles DFT calculations, namely $170$ \nm \cite{Mannix2015} and $163$ \nm \cite{Mortazavi2016b} for elongation along the zigzag direction and $398$ \nm \cite{Mannix2015} and $382$ \nm \cite{Mortazavi2016b} when stretched along the armchair direction.
The lower Young's modulus and higher fracture strain in the zigzag direction in comparison to the armchair one are due to the  buckling along the zigzag direction of borophene, shown in figure \ref{fig1}. 
The agreement of our results for the 2D Young's modulus with first principles calculations reveals that the utilized ReaxFF potential can accurately describe the atomic interaction at low strain levels within the elastic region, even if it does not yield the same lattice parameters as first principles calculations. This observation will also be valid for our thermal conductivity evaluation in the next few paragraphs. 

The behavior of the strees-strain curves shown in figure \ref{fig2} are quite different. Notably there seem to be two distinct linear regimes when the sample is stretched along the zigzag direction. We atribute this behavior to the buckling of the structure along the zigzag direction. In the first linear region the buckling of borophene is decreased due to the strain, and the sample becomes flatter. In the second regime the now in-plane B-B bonds are stretched up to their rupture point.

\begin{figure}[htb]
\centering
\includegraphics[width=\linewidth]{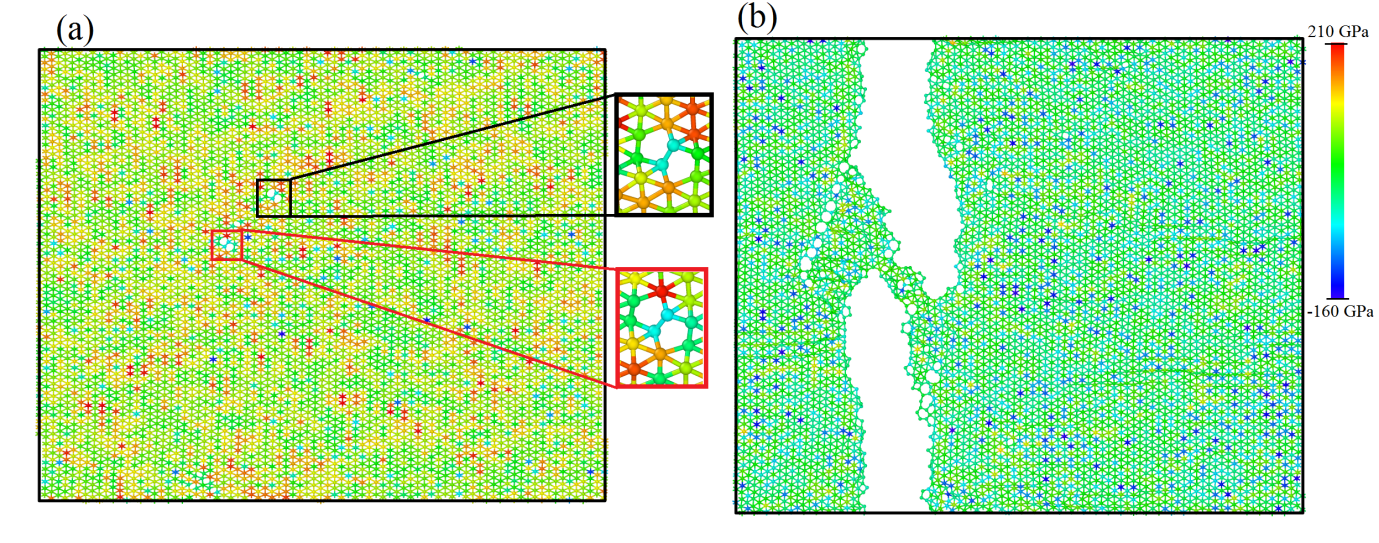}
\caption{Uniaxial deformation process of single-layer borophene stretched along the zigzag direction (a) shortly before the rupture and (b) at failure point. The stress values are the uniaxial stress distribution along the loading direction. The OVITO package was used for the illustration of this figure \cite{Stukowski2010}.}
\label{fig3}
\end{figure}

In figure \ref{fig3}, the deformation of a single-layer corrugated borophene sheet stretched along its zigzag direction is depicted. For borophene sheets under uniaxial tensile loading along the zigzag and armchair directions, we found that the structures extend uniformly and remain defect-free up to strain levels close to rupture. For a borophene membrane stretched along the zigzag direction, shortly before rupture the first B-B bond breakages occur (figure \ref{fig3}(a)), resulting in the formation of pentagons (figure \ref{fig3}(a) insets). Instantly after the formation of these initial pentagon defects the ultimate tensile strength is reached, and the coalescence of defects happen, resulting in the formation of a crack that grows rapidly perpendicular to the loading direction and leads to the sample rupture (figure \ref{fig3}(b)). Based on our reactive atomistic modeling, corrugated borophene presents a brittle failure mechanism at room temperature. This conclusion is corroborated by the fact that the initial defect formation and sample rupture occur at very close strain levels.   

Non-equilibrium molecular dynamics simulations for borophene sheets were performed along armchair and zigzag directions,  for samples of increasing length in order to assess possible size effects on the thermal conductivity. In NEMD simulations the calculated lattice thermal conductivity presents a strong size dependency when the sample length is smaller than the phonon mean-free path of the structure \cite{Schelling2002,Xu2014,Fugallo2014,Zhu2014,Zhang2015,Barbarino2015,Neogi2015,Majee2016,Mortazavi2016a,Pereira2016}. Accordingly, the thermal conductivity of borophene sheets obtained in our simulations increased with the sample length, $L$. As a common approach, we can express the length dependence of the thermal conductivity with the ballistic to diffusive transition equation
\begin{equation}
\kappa(L) = \frac{\kappa}{1+\Lambda_{eff}/L},
\label{eq:fit}
\end{equation}
where $\kappa$ is the intrinsic thermal conductivity of an infinite-sized sample, and $\Lambda_{eff}$ is an effective phonon mean free path for the material \cite{Neogi2015,Mortazavi2016a,Pereira2016}. Notice that from this expression we have $\kappa(L)=\kappa/2$ when $L=\Lambda_{eff}$. Therefore, adjusting the above equation to the data points obtained from NEMD simulations with different lengths we can determine both free-parameters, $\kappa$ and $\Lambda_{eff}$.

\begin{figure}[htb]
\centering
\includegraphics[width=\linewidth]{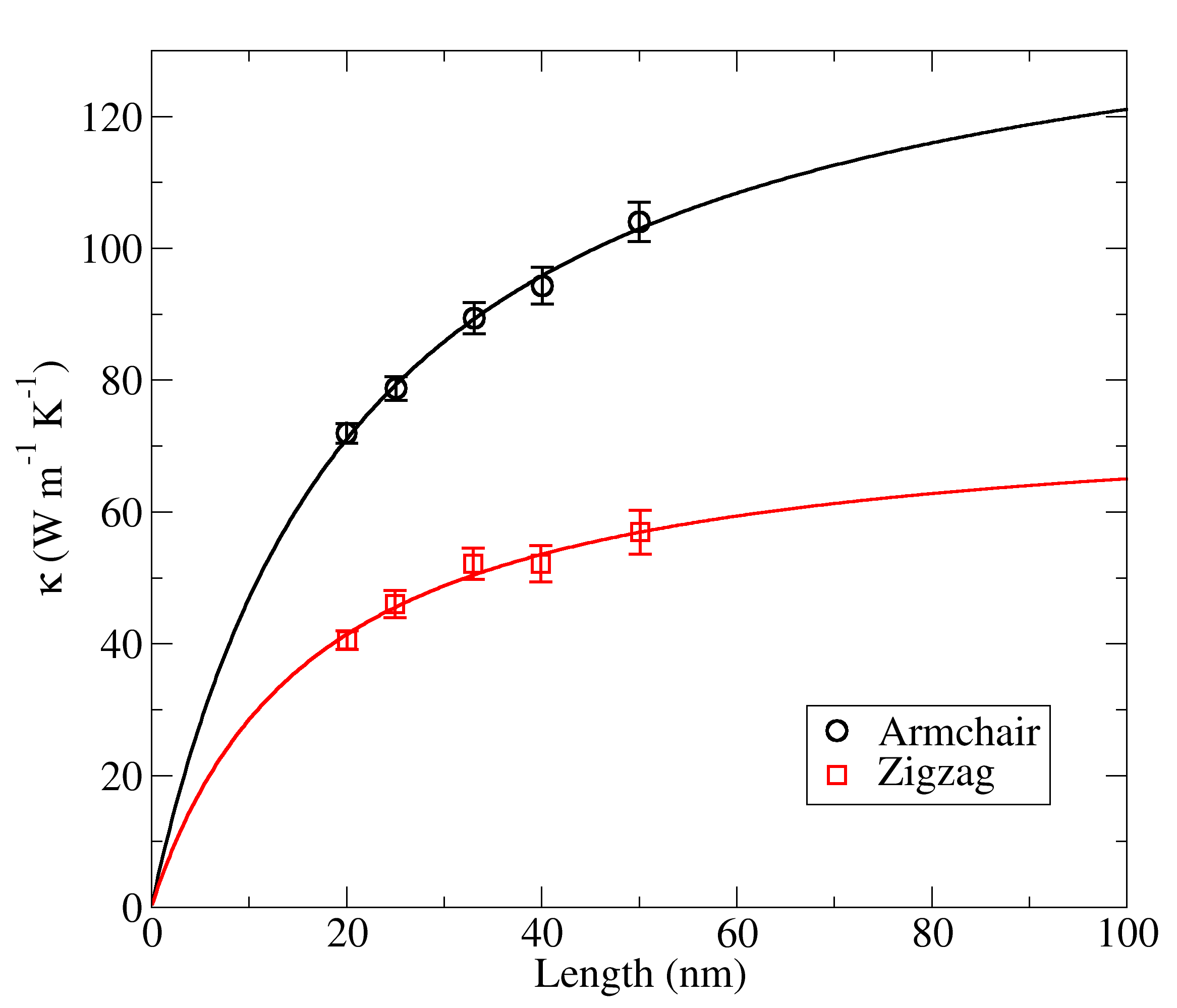}
\caption{Thermal conductivity of borophene as a function of length along armchair and zigzag directions. Data points from NEMD simulations and lines from equation \ref{eq:fit}. Thermal conductivities are $75.9 \pm 5.0$ \wmk and $147 \pm 7.3$ \wmk along zigzag and armchair directions, respectively.}
\label{fig4}
\end{figure}

Figure \ref{fig4} shows the data points calculated from NEMD simulations at $300$ K, and the solid lines indicate the corresponding fit to the data, from which we calculate conductivities of $75.9 \pm 5.0$ \wmk and $147 \pm 7.3$ \wmk along zigzag and armchair directions respectively. Meanwhile, the effective phonon mean free path take values of $16.7 \pm 1.7$ nm and $21.4 \pm 1.0$ nm for the zigzag and armchair directions. The data shows a clear anisotropy in the conduction of heat along the in-plane directions of borophene, with the conductivity along the armchair direction being about twice as large as in the zigzag direction. Such feature could be explored in the construction of future phononic devices \cite{Wagner2016}. 

\begin{figure}[htb]
\centering
\includegraphics[width=\linewidth]{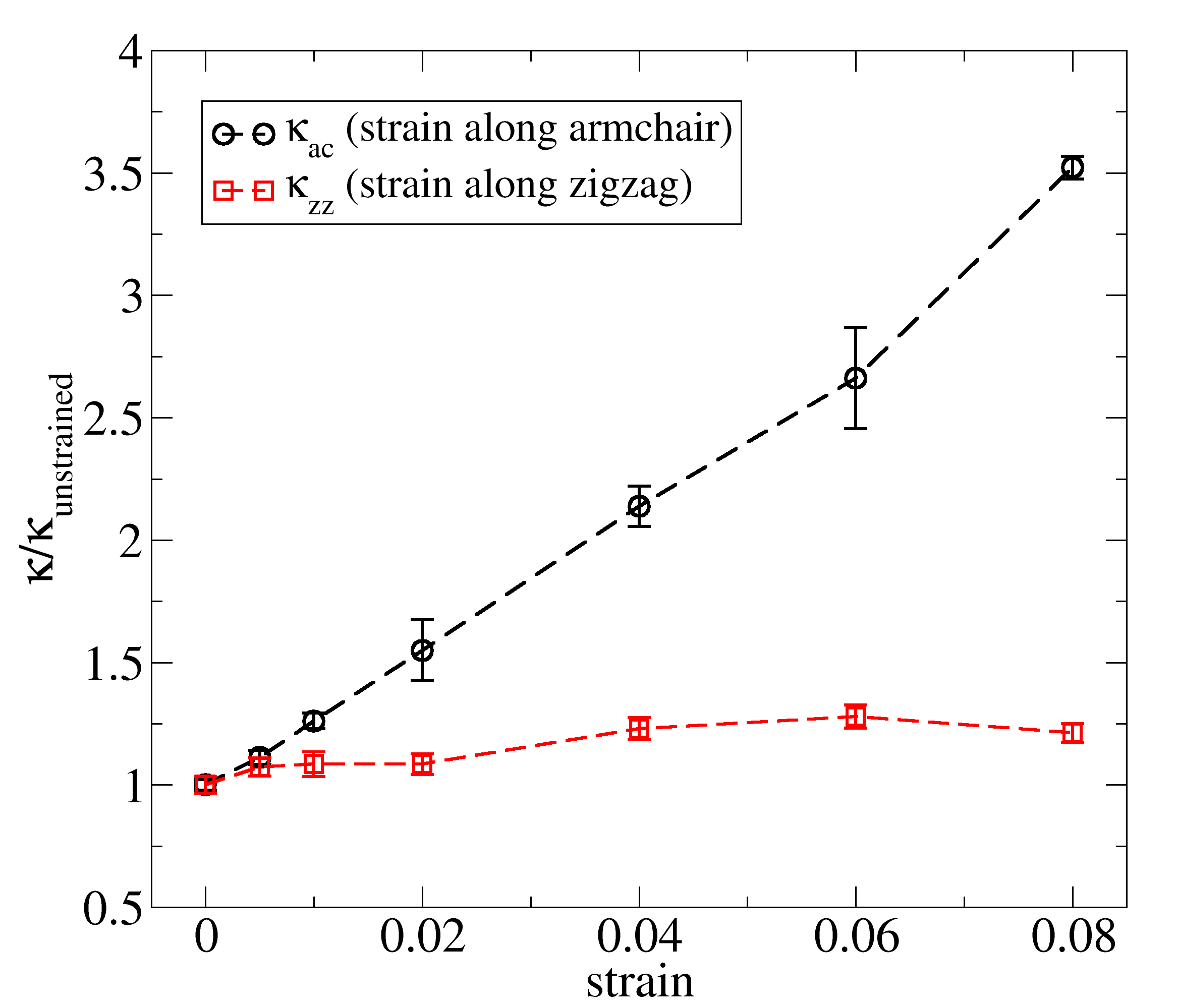}
\caption{Normalized thermal conductivity of borophene under strain. Notice the significant increase when the sample is strained along the armchair direction.}
\label{fig5}
\end{figure}

{Finally, we investigated the effect of mechanical strain on the the thermal conductivity of borophene along each direction independently, as shown in figure \ref{fig5}. When the sample is strained along the armchair direction there is a significant increase in thermal conductivity along that direction. Meanwhile, when the sample is strained along the zigzag direction there is a much smaller increase in thermal conductivity along that direction. For a strain of 8\% along the armchair direction the thermal conductivity increases by a factor of 3.5 (~250\%), whereas for the same amount of strain along the zigzag direction the increase is only by a factor of 1.2 (~20\%).
This feature reinforces the prospective application of borophene in the construction of  phononic devices \cite{Wagner2016}. }

\section{Discussion}

In the development of this study, several interatomic potentials have been tested to describe the structure of borophene, including the Tersoff potential \cite{Tersoff1988,Tersoff1988a,Matsunaga2000,Mortazavi2015a}, and different versions of ReaxFF.
We have found that the ReaxFF parameter set developed by Weismiller et al. \cite{Weismiller2010} yields the closest predictions to first-principles results. 
Even though the 2D Young's modulus predicted by reative MD are within $10$\% of first principles results, and the observed trends being in accordance with previous DFT results, the tensile strength and fracture strain are both much larger than predicted by DFT calculations \cite{Mortazavi2016b}.
In fact, along the zigzag direction, the failure strain of borophene ($\sim 25$\%) is close to that of graphene ($20$\%-$27$\%) \cite{Lee2008,Le2016a,Mortazavi2016}. 
We believe this overestimation in tensile strength and failure strain could probably be decreased by adjusting the cutoff length for B-B bonds in the ReaxFF parameter set.
One should also consider that according to DFT calculations the equilibrium B-B bond lengths are not equal for the two different bonds in corrugated borophene \cite{Mannix2015,Mortazavi2016b}, whereas in ReaxFF, all B-B bonds are treated using a single set of parameters. 
Nonetheless, we atribute the anisotropy in 2D Young's modulus and fracture strain to the buckling along the zigzag direction of borophene. In our interpretation, this buckling is also responsible for the appearance of two linear regimes in the strees-strain curve along the zigzag direction. Where the first part is due to the flattening of the borophene sheet and the second is due to the stretching of in-plane bonds.

\begin{figure}[htb]
\centering
\includegraphics[width=\linewidth]{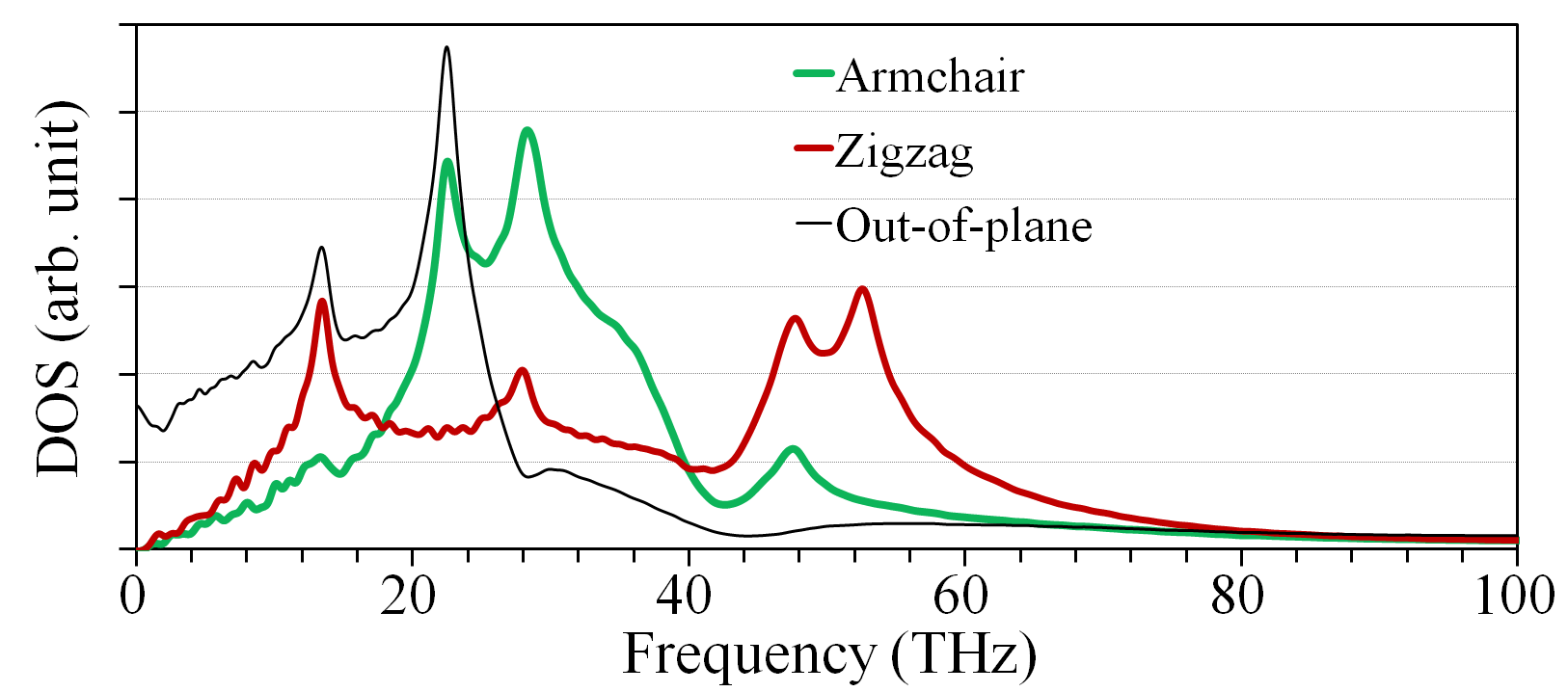}
\caption{Calculated vibrational density of states (DOS) for pristine borophene along different directions.}
\label{fig6}
\end{figure}

The observed anisotropy in the lattice thermal conductivity is consistent with the results for the elastic moduli discussed above, where the elastic constant along the armchair direction was found to be larger than along the zigzag direction by a factor of $2$.
In order to complement our understanding of the underlying mechanism resulting in the anisotropic thermal conductivity of borophene, we calculated the phonon density of states (DOS) as  the Fourier transform of the velocity autocorrelation function.
Figure \ref{fig6} shows the vibrational density of states projected onto the in-plane directions, as well as the out-of-plane direction. 
Out-of-plane phonons have the largest DOS in the low-frequency regime (below $20$ THz), being most likely the main heat carriers in borophene. 
Nonetheless,  in the low-frequency region up to $16$ THz we observe a larger DOS for vibrations projected onto the zigzag direction.
We also observe a peak around $14$ THz for vibrations along the zigzag direction, which is caused by flat bands in the phonon dispersion and yield lower phonon group velocities along that direction. 
Finally, since there are more phonon modes for scattering along the zigzag direction, it could also shorten the mean-free-path and phonon lifetimes along that direction.

Our simulation results are remarkably close to the prediction presented in a very recent first-principles based study of the thermal properties of borophene \cite{Sun2016}. In their work, Sun et al. did not assume a thickness for borophene sheets, but if we scale their results with the nominal thickness used in our simulations we obtain $\kappa_{zigzag} \approx 72$ \wmk and $\kappa_{armchair} \approx 145$ \wmk. Thus a factor of $2$ anisotropy is observed in the in-plane thermal conductivities, which is attributed by them to the larger phonon group velocities along the armchair direction. 
This observation is consistent with the smaller variation we found in the effective phonon mean free path along armchair and zigzag directions.
The difference in group velocities reported by Sun et al., along with our DOS analysis provides a consistent explanation for the predicted anisotropy in the in-plane thermal conductivity of borophene \cite{Mortazavi2016,Pereira2016}.
However, since their calculations are based on the quasi-harmonic approximation, which considers only three-phonon scattering processes, one could expect their conductivity to be larger than ours, which accounts for all possible anharmonic scattering processes. In fact, a comparison between their first-principles calculations and our MD results leads to the conclusion  that scattering events including more than three phonons do not have a significant impact in the thermal conductivity of borophene.

\section{Summary}

We performed reactive molecular dynamics simulations to explore the mechanical response and the lattice thermal conductivity of free-standing corrugated borophene at room temperature.
The ReaxFF potential developed by Weismiller et al. was used to model atomic interactions in borophene sheets. 
Although the ReaxFF potential overestimate the in-plane lattice parameters of borophene, we find that it predicts the elastic modulus and thermal conductivity of this novel material in excellent agreement with first-principles calculations, with a much lower computational cost.
According to our reactive molecular dynamics simulations, the 2D Young's modulus of borophene was estimated to be $188$ \nm and $403$ \nm along  zigzag and armchair directions, respectively, which are within $10$\% of first-principles predictions. 
This anisotropy in elastic moduli is attributed to the buckling of the borophene crystal structure along the zigzag direction.
We also performed non-equilibrium molecular dynamics simulations to calculate the lattice thermal conductivity of borophene sheets of increasing length 
The intrinsic thermal conductivity of an infinite borophene sheet, as well as an effective phonon mean free path were obtained from the ballistic to diffusive transition equation. At room temperature, the thermal conductivity of borophene along its zigzag and armchair directions were predicted to be $75.9 \pm 5.0$ \wmk and $147 \pm 7.3$ \wmk, respectively. 
Meanwhile, the effective phonon mean free path take values of $16.7 \pm 1.7$ nm and $21.4 \pm 1.0$ nm for the zigzag and armchair directions.
In this case, the anisotropy is attributed to differences in the density of states of low-frequency phonons, which correlate to lower group velocities and possibly shortens phonon lifetimes along the zigzag direction.
{We also observe that when borophene is strained along the armchair direction there is a significant increase in thermal conductivity along that direction. Meanwhile, when the sample is strained along the zigzag direction there is a much smaller increase in thermal conductivity along that direction. For a strain of 8\% along the armchair direction the thermal conductivity increases by a factor of 3.5 (~250\%), whereas for the same amount of strain along the zigzag direction the increase is only by a factor of 1.2 (~20\%).}
Interestingly, the anisotropy ratio for the thermal conductivity and elastic modulus were found to be almost the same, and also consistent with recent first-principles calculations. 
Furthermore, comparing our phonon thermal conductivity predictions with first-principles calculations based on the quasi-harmonic approximation, which considers only three-phonon scattering processes, indicates that scattering events including more than three phonons do not have a significant impact in the thermal conductivity of borophene.
Our predictions for the elastic moduli and lattice thermal conductivity are in agreement with recent first principles results, at a fraction of the computational cost.
We expect our simulations to serve as a guide for future experiments concerning the mechanical and thermal properties of borophene and related novel 2D materials.

\section{Acknowledgements}
The authors would like to thank L.D. Machado for a critical reading of the manuscript. BM and TR greatly acknowledge the financial support by European Research Council for COMBAT project (Grant No. 615132). MQL   was supported by Vietnam National Foundation for Science and Technology Development (NAFOSTED) under the grant number: 107-02-2017-02. LFCP acknowledges financial support from the Brazilian government agency CAPES for the project ``Physical properties of nanostructured materials'' (Grant No. 3195/2014) via its Science Without Borders program and provision of computational resources by the High Performance Computing Center (NPAD) at UFRN.



\end{document}